\documentclass{ifacconf}
\usepackage[utf8]{inputenc}
\usepackage[english]{babel}
\newtheorem{theorem}{Theorem}

\usepackage{blindtext}
\usepackage{graphicx}
\usepackage[utf8]{inputenc}
\usepackage{graphics} 
\usepackage{epsfig} 
\usepackage{mathptmx}
\usepackage{amsmath} 
\usepackage{amssymb}  
\usepackage{multicol}
\usepackage{mathtools, cuted}
\usepackage[cmintegrals]{newtxmath}
\usepackage{epstopdf}
\usepackage{graphics} 
\usepackage{epsfig} 
\usepackage{mathptmx} 
\usepackage{amsmath} 
\usepackage{amssymb}  
\usepackage{multicol}
\usepackage{mathtools, cuted}
\usepackage[cmintegrals]{newtxmath}
\usepackage{mathtools}

\usepackage{mathptmx}
\usepackage{helvet}
\usepackage{courier}
\usepackage{cancel}
\usepackage{subfigure}
\usepackage{type1cm}
\usepackage{epsfig} 
\usepackage{mathptmx} 
\usepackage{amsmath} 
\usepackage{amssymb}  
\usepackage{multicol}
\usepackage{mathtools, cuted}
\usepackage[cmintegrals]{newtxmath}
\usepackage{makeidx}         
\usepackage{algorithmic}
\DeclareMathOperator*{\argmin}{\arg\!\min}

\usepackage{multicol}        
\usepackage{nomencl}
\usepackage[usenames, dvipsnames]{color}

\usepackage{graphicx}      
\usepackage{natbib}        
\begin{document}
\begin{frontmatter}

\title{An Integrative Data-Driven Physics-Inspired Approach to Traffic Congestion Control\thanksref{footnoteinfo}} 

\thanks[footnoteinfo]{This work has been supported by the National Science Foundation under Award Nos. 1739525 and 1914581.}

\author[First]{Hossein Rastgoftar} 
\author[Second]{Ella Atkins}

\address[First]{Department
of Aerospace Engineering, University of Michigan, Ann Arbor,
MI, 48109 USA, (e-mail: hosseinr@umich.edu)}
\address[Second]{Department
 of Aerospace Engineering, University of Michigan, Ann Arbor,
 MI, 48109 USA, (e-mail: ematkins@umich.edu)}

\begin{abstract}                

This paper offers an integrative data-driven physics-inspired approach to model and control traffic congestion in a resilient and efficient manner. While existing physics-based approaches commonly assign density and flow traffic states by using the Fundamental Diagram, this paper specifies the flow-density relation using past traffic information recorded in a time sliding window with a constant horizon length. With this approach, traffic coordination trends can be consistently learned and incorporated into traffic planning. This paper also models traffic coordination as a probabilistic process and obtains traffic feasibility conditions using linear temporal logic. Model productive control (MPC) is applied to control traffic congestion through the boundary of the traffic network. Therefore, the optimal boundary inflow is assigned as the solution of a constrained quadratic programming problem. 
\end{abstract}


\end{frontmatter}

\section{Introduction}
Urban traffic congestion management has been an active research area, and physics-based modeling of traffic coordination has been extensively studied by researchers over the past three decades. 
 It is common to spatially discretize a network of interconnected roads (NOIR) using the Cell Transmission Model (CTM) which applies mass conservation to model traffic coordination [\citep{daganzo1995cell, gomes2006optimal}]. To control and analyze traffic congestion, the Fundamental Diagram is commonly applied to assign a flow-density relation at every traffic cell. While the Fundamental Diagram can successfully determine the traffic state for small-scale urban road networks, it may not properly function for congestion analysis and control in large traffic networks. Modeling of backward propagation, spill-back congestion, and shock-wave propagation is quite challenging. The main objective of this paper is to deal with the aforementioned challenges of modeling and control of traffic congestion modeling and control.  In particular, this paper contributes a novel integrative data-driven physics-inspired approach to \textit{obtain a microscopic data-driven traffic coordination model} and \textit{resiliently control congestion in large-scale traffic networks}.

Researchers have proposed light-based  and physics-based control approaches to address traffic coordination challenges. Fixed-cycle control is the traditional approach for the operation of traffic signals at intersections. The traffic network study tool
[\citep{robertson1969transyt, tiwari2008continuity}] is a standard fixed-cycle control tool for optimization of the signal timing.  Refs. [\citep{balaji2011type, chiu1992adaptive}] offer fuzzy-based signal control approaches to optimize the green time interval at junctions.  Physics-based traffic coordination approaches commonly use the Fundamental Diagram to determine traffic state (flow-density relation) [\citep{zhang2012ordering, zhang2011transitions}], model dynamic traffic coordination [\citep{han2012continuous}], incorporate spillback congestion [\citep{gentile2007spillback, adamo1999modelling}] and backward propagation [\citep{gentile2015using, long2008urban}] effects into traffic simulation, or specify the feasibility conditions for traffic congestion control. 
Ref. [\cite{jafari2018structural}] integrate first order traffic dynamics inspired by mass flow conservation, dynamic traffic  assignment [\citep{peeta2001foundations, janson1991dynamic}], and a cell transmission model [\citep{daganzo1995cell, daganzo1994cell}] to model and control freeway traffic coordination.  Model predictive control (MPC) is the common control approach for model-based traffic coordination optimization [\citep{lin2012efficient, jamshidnejad2018sustainable, tettamanti2014robust}]. Ref. \cite{baskar2012traffic} applies MPC to determine the optimal platooning speed for automated highway systems (AHS). Furthermore, researchers have applied fuzzy logic [\citep{kammoun2014adapt, collotta2015novel, pau2018smart, yusupbekov2016adaptive}], neural network \citep{moretti2015urban, tang2017improved, akhter2016neural, kumar2015short}, Markov Decision Process (MDP) [\citep{ong2016markov, haijema2008mdp}], formal methods \citep{coogan2017formal, coogan2015traffic} and mixed nonlinear programming (MNLP) [\citep{christofa2013person}] for model-based traffic management.  Optimal control [\citep{ jafari2018structural, wang2018dynamic}] approaches have also been proposed.

This paper offers a new data-driven approach for control traffic coordination in a network of interconnected roads (NOIR). We model traffic coordination as a mass conservation problem governed by the continuity partial differential equation (PDE). By spatial and temporal discretization of traffic coordination, traffic dynamics is expressed by a probabilistic process controlled through the boundary road elements of the NOIR. The paper uses linear temporal logic to formally specify the feasibility conditions at NOIR road elements. Given traffic feasibility conditions, optimal boundary inflow is assigned as the solution of an adaptive model-predictive control with parameters that are consistently learned based on the empirical traffic information. Therefore, the optimal boundary inflow is continuously assigned as the solution of a constrained quadratic programming problem and incorporated into planning. 

This paper is organized as follows. Preliminary notions of graph theory presented in Section \ref{Graph Theory Notions} are followed by a problem statement in Section \ref{Problem Statement}. Section \ref{Traffic Coordination Modeling} models traffic coordination as a mass-conservation problem followed by traffic congestion boundary control in Section \ref{Traffic Coordination Control}. Simulation results presented in Section \ref{Simulation Results} are followed by concluding remarks in Section \ref{Conclusion}.
\section{Graph Theory Notions}
\label{Graph Theory Notions}
NOIR roads are filled by a finite number of serially-connected road elements, where $i\in \mathcal{V}$ represents a unique road element. The node set $\mathcal{V}$ can be expressed as
\[
\mathcal{V}=\mathcal{V}_{in}\bigcup \mathcal{V}_{out}\bigcup \mathcal{V}_I,
\]
where $\mathcal{V}_{in}=\{1,\cdots,N_{in}\}$, $\mathcal{V}_{out}=\{N_{in}+1,\cdots,{N_{out}}\}$, and $\mathcal{V}_{I}=\{{N_{out}+1},\cdots,N\}$ define index numbers of inlet, outlet, and interior road elements, respectively. Interaction between road elements are defined by graph $\mathcal{G}\left(\mathcal{V},\mathcal{E}\right)$, where $\mathcal{E}\subset \mathcal{V}\times \mathcal{V}$ specify edges of graph $\mathcal{G}$. For every road element $i\in \mathcal{V}$, an \textit{in-neighbor set} $\mathcal{I}_{i}$ specifies upstream adjacent road elements and \textit{out-neighbor set} $\mathcal{O}_{i}$ defines downstream adjacent road elements. Traffic enters $i\in \mathcal{V}$ from an in-neighbor node belonging to $\mathcal{I}_{i}$ and exits from $i\in \mathcal{V}$ toward an out-neighbor node belonging to $\mathcal{O}_{i}$. 


\section{Problem Statement}
\label{Problem Statement}
The traffic coordination control problem is defined by the finite state abstraction $\mathrm{M}$ given by tuple
\[
\mathrm{M}=\left(\mathbf{X},\mathbf{U},\mathcal{A},\mathcal{F},\mathcal{P},\mathrm{C}\right),
\]
where $\mathbf{X}\subset \mathbb{R}^{N-N_{out}}$ is a finite subspace of $\mathbb{R}^{N-N_{out}}$, vector $\mathbf{x}=\left[\rho_{N_{out}+1}~\cdots~\rho_N\right]^T\in \mathbf{X}$ defines traffic density of interior road elements across the NOIR, where $\rho_i$ is the traffic density at interior road element $i\in \mathcal{V}_I$. $\mathbf{U}\subset \mathbb{R}^{N_{in}}$ is a finite subspace of $\mathbb{R}^{N_{in}}$, vector $\mathbf{u}=\left[u_{1}~\cdots~u_{N_{in}}\right]^T\in \mathbf{U}$ defines the boundary inflow, where $u_i$ is the inflow at inlet boundary element $i\in \mathcal{V}_{in}$. 
Furthermore, $\mathcal{F}:\mathbf{X}\times \mathbf{U}\rightarrow \mathcal{X}$ is the \textit{traffic state transition function} defined as follows:
\begin{equation}
\label{transfun}
    \mathcal{F}\left(\mathbf{x},\mathbf{u},a\right)=\mathbf{A}_{a}\mathbf{x}+\mathbf{B}\mathbf{u}
\end{equation}
 where $a=\in \mathcal{A}$ is a discrete action characterizing traffic tendency at every discrete time $k$. Note that $a\in \mathcal{A}$ is consistently learned and incorporated into planning. Also, $\mathbf{B}\in \mathbb{R}^{\left(N-N_{out}\right)\times N_{in}}$ is constant, and $\mathbf{A}_a:\mathcal{A}\rightarrow \mathbb{R}^{\left(N-N_{out}\right)\times \left(N-N_{out}\right)}$ is the \textit{tendency matrix} defined in Section \ref{Traffic Coordination Modeling}. 
 Moreover, $\mathcal{P}$ is the set of atomic propositions that are used to provide traffic feasibility conditions. $\mathrm{C}:\mathcal{X}\times\mathbf{U}\times \mathcal{A}\rightarrow \mathbb{R}_{\geq 0}$ is the traffic coordination cost defined based on the traffic density distribution across the NOIR.

\section{Traffic Coordination Modeling}
\label{Traffic Coordination Modeling}
The mass-conservation law is applied to obtain microscopic traffic dynamics  across the NOIR. Therefore, traffic coordination at road element $i\in \mathcal{V}_I$ is given by
\begin{equation}
\label{main}
    \rho_i[k+1]=\rho_i[k]+y_{i,a[k]}[k]-z_{i,a[k]}[k],
\end{equation}
where $k\in \mathbb{Z}$ denotes discrete time, $\rho_i[k]$ is the traffic density of road element $i$,  $a[k]\in \mathcal{A}$ is a tendency action executed over the time interval $t\in [t_k,t_{k+1}]$, 
\begin{subequations}
\begin{equation}
   y_{i,a[k]}[k]=
    \begin{cases}
    u_{i,k}&i\in \mathcal{V}_{in}\\
    \bar{q}_{_{i,j,a[k]}}\bar{p}_{_{j,a[k]}}\rho_i[k]&i\in \mathcal{V}\setminus \mathcal{V}_{in}
    \end{cases}
    ,
\end{equation}
\begin{equation}
    z_{i,a[k]}[k]=
    \bar{p}_{i,a[k]}\rho_i[k]
\end{equation}
\end{subequations}
 are the traffic inflow and outflow respectively at road element $i$ over  time interval $[t_k,t_{k+1}]$. Note that  $\bar{p}_{_{i,a[k]}}\in [0,1]$ is the outflow probability of road element $i\in \mathcal{V}_I$, determining the fraction of cars leaving road element $i\in \mathcal{V}_I$ over time interval $t\in [t_k,t_{k+1}]$. Also, tendency probability $\bar{q}_{_{j,i,a[k]}}$ is the fraction of $z_{_{i,a[k]}}[k]$ directed from $i\in \mathcal{V}\setminus \mathcal{V}_{out}$ toward $j\in \mathcal{O}_{i}$ at time $t\in [t_k,t_{k+1}]$, where
 \[
 \sum_{j\in \mathcal{O}_{i}}\bar{q}_{j,i,a[k]}=1.
 \]

\subsection{Traffic state transition function:}  
Over time-interval $[t_k,t_{k+1}]$, we define positive-definite and diagonal matrix 
\begin{equation}
    \mathbf{P}_{a[k]}=
    \begin{bmatrix}
    \bar{p}_{_{N_{out}+1,a[k]}}&&0\\
    &\ddots&\\
    0&& \bar{p}_{_{N,a[k]}}
    \end{bmatrix}
    \in \mathbb{R}^{\left(N-N_{out}\right)\times \left(N-N_{out}\right)},
\end{equation}
where $a[k]\in \mathcal{A}$. We also define non-negative matrix 
$\mathbf{Q}_{a[k]}=\left[{\bar{q}_{{a[k]}_{ij}}}\right]=\left[\bar{q}_{i+N_{out},j+N_{out},a[k]}\right]\in \mathbb{R}^{\left(N-N_{out}\right)\times \left(N-N_{out}\right)}$ with $ij$ entry ${\bar{q}_{{a[k]}_{ij}}}=\bar{q}_{i+N_{out},j+N_{out}[k],a[k]}$ specifying the tendency of traffic at interior node $j+N_{out}\in \mathcal{V}_I$ to move towards $\left(i+N_{out}\right)\in \mathcal{O}_{j+N_{out}[k]}$ at any time $t\in [t_k,t_{k+1}]$.  

\textbf{Traffic Tendency Matrix:} Given $\mathbf{P}_{a[k]}$ and $\mathbf{Q}_{a[k]}$, we define the \textit{tendency matrix} $\mathbf{A}_{a[k]}\in \mathbb{R}^{\left(N-N_{out}\right)\times \left(N-N_{out}\right)}$ as follows:
\begin{equation}
\label{alambdatheta}
a[k]\in \mathcal{A},\qquad     \mathbf{A}_{a[k]}=\mathbf{I}-\mathbf{P}_{a[k]}+\mathbf{Q}_{a[k]}\mathbf{P}_{a[k]},
\end{equation}
where $\mathbf{I}\in \mathbb{R}^{\left(N-N_{out}\right)\times \left(N-N_{out}\right)}$ is the identity matrix. Assuming traffic is updated by Eq. \eqref{main} at every road element $i\in \mathcal{V}_I$, density vector $\mathbf{x}[k]=\left[\rho_{N_{out}+1}[k]~\cdots~\rho_{N}[k]\right]^T$ is updated by the following network dynamics:
\begin{equation}
\label{MAIN}
a[k]\in \mathcal{A},\qquad 
    \mathbf{x}[k+1]=\mathcal{F}\left(\mathbf{x}[k],\mathbf{u}[k],a[k]\right)=\mathbf{A}_{a[k]}\mathbf{x}[k]+\mathbf{B}\mathbf{u}[k],
\end{equation}
where $\mathbf{u}[k]=\left[u_{1}[k]~\cdots~u_{N_{in}}[k]\right]^T$ is the boundary inflow vector,   $\mathbf{B}=\left[B_{ij}\right]\in \mathbb{R}^{\left(N-N_{out}\right)\times N_{in}}$, and $B_{ij}$ is a constant matrix with $ij$ entry
\begin{equation}
    B_{ij}=
    \begin{cases}
    1&j\in \mathcal{I}_{i+N_{out}}\\
    0&\mathrm{otherwise}
    \end{cases}
    .
\end{equation}
\begin{theorem}
The traffic state transition, defined by dynamics \eqref{MAIN}, is BIBO stable when the following conditions are satisfied:
\begin{enumerate}
    \item{There exists at least one directed path from every inlet boundary road element toward the interior of road element $i\in \mathcal{V}_I$.} 
    \item{There exists at least one directed path from the interior of road element $i\in \mathcal{V}_I$ toward every outlet boundary road element.} 
\end{enumerate}
\end{theorem}
\textbf{Proof:}  
Because there exists a path from each boundary node to every interior node of graph {\color{black}$\mathcal{G}$}, matrix $\mathbf{A}_{a[k]}$ is irreducible, and the sum of the column entries of $\mathbf{A}_{a[k]}$ is either negative or zero. Entries of column $i$ of matrix $\mathbf{A}_{a[k]}$ sum to a negative number between $-1$ and $0$ if no out-neighbors of road element $i+N_{out}$ are outlet boundary nodes, i.e. $\mathcal{O}_{i+N_{out}}\bigcap \mathcal{V}_{out}=\emptyset$. Otherwise, the sum of the entries of column $i$ of matrix  $\mathbf{A}_{a[k]}$ is $0$. Consequently, the spectral radius of $\mathbf{A}_{a[k]}$ denoted by $r_{a[k]}$ is less than $1$. Therefore, eigenvalues of $\mathbf{A}_{a[k]}$ are all located inside a disk with radius $r_{a[k]}<1$  {\color{black}$[t_k,t_{k+1}]$} with a center positioned at the origin. 

When $\mathbf{x}[k]$ is updated by discrete traffic dynamics \eqref{MAIN}, we can write
\begin{equation}
\label{MAINNNNTheorem4}
\begin{split}
 a[k]\in \mathcal{A},\qquad   \mathbf{x}[k+1]=&\mathbf{\Gamma}_{k}\mathbf{x}[1]+\mathbf{\Omega}_{k}\mathbf{B}\begin{bmatrix}
    \mathbf{u}[1]\\
    \vdots\\
    \mathbf{u}[k]
    \end{bmatrix}
   \end{split}
,
\end{equation}
where 
\begin{subequations}
\begin{equation}
\mathbf{\Gamma}_k=\mathbf{A}_{a[k]}\times\mathbf{A}_{a[k-1]}\times  \cdots\mathbf{A}_{a[1]}\in \mathbb{R}^{\left(N-N_{out}\right)\times\left(N-N_{out}\right)},
\end{equation}
\begin{equation}
\mathbf{\Omega}_k=
    \begin{bmatrix}
    \mathbf{A}_{a[k]}^{k-1}&\cdots&\mathbf{A}_{a[1]}&\mathbf{I}
    \end{bmatrix}
    \in \mathbb{R}^{\left(N-N_{out}\right)\times k\left(N-N_{out}\right)},
\end{equation}
\end{subequations}
where $a[1],\cdots,a[k-1],a[k]\in \mathcal{A}$.      
Because eigenvalues of matrix ${\mathbf{A}}_{a[h]}$ ($h\in \{1,\cdots,k\}$) are all placed inside a disk with spectral radius $r_{a[h]}<1$ centered at the origin, 
\begin{equation}
    \|\mathbf{\Gamma}_k\mathbf{x}[1]\|\leq\|\mathbf{x}[1]\| \prod_{h=1}^kr_{a[h]}\leq \|\mathbf{X}[1]\|,
\end{equation}
and
\begin{equation}
a[h]\in \mathcal{A},\qquad    \|\mathbf{A}_{a[k]}^h\mathbf{B}\mathbf{u}[k]\|\leq r_{a[h]}\|\mathbf{B}\mathbf{u}[k]\|\leq \|\mathbf{B}\mathbf{u}[k]\|
\end{equation}
is bounded assuming input vector $\mathbf{u}[k]$ is bounded at any discrete time $k$. Consequently, both terms are bounded on the right hand side of Eq. \eqref{MAINNNNTheorem4} and the BIBO stability of the traffic dynamics \eqref{MAINNNNTheorem4} is proven for every $a\in \mathcal{A}$.

\subsection{Learning of Traffic Tendency}  
Let human intent be defined by discrete set
\[
\mathcal{A}=\{a_l=\left(a_{N_{out}+1,l},\cdots,a_{i,l},\cdots,a_{N,l}\right)\big|i\in \mathcal{V}_I,~l=1,\cdots,n_a\}
\]
where $n_a=\left|\mathcal{A}\right|$ is the cardinality of set $\mathcal{A}$ and $a_{l}$ characterizes the $l$-th possible traffic tendency action.  It is assumed that 
\begin{subequations}
\begin{equation}
    \mathcal{P}=\left\{\left(\bar{p}_{_{N_{out}+1},a},\cdots,\bar{p}_{_{N,a}}\right)\big|a\in \mathcal{A}\right\},
\end{equation}
\begin{equation}
    \mathcal{W}=\left\{\left(\mathbf{q}_{N_{out+1},a},\cdots,\mathbf{q}_{N,a}\right)\bigg|a\in \mathcal{A},~\mathbf{q}_i=\bigcup_{j\in \mathcal{O}_{N_{Out}+1}}q_{_{j,N_{out}+1,a}}\right\}
\end{equation}
\end{subequations}
are finite sets specifying all possible values for outflow and tendency probabilities for every discrete action $a\in \mathcal{A}$. It is further assumed that traffic information data including traffic density $\rho_i$ and traffic outflow $z_{i}$ are available at discrete times $k-1$, $\cdots$, $k-L$ ($h=1,\cdots,k$) for every discrete time $k\in \mathbb{Z}$, where $L$ is the length of the time sliding window recording traffic history information. Therefore, cost function
\[
\mathcal{C}_{H}(a[k])=\sum_{h=1}^L\sum_{i\in \mathcal{V}_I}\left(z_i[k-h]-\sum_{j\in \mathcal{O}_{i[k-h]}}p_{i,a[k-h]}q_{j,i,a[k-h]}\right)^2,
\]
is known at every discrete time $k$, and human intent $\theta[k]$ is obtained as follows:
\begin{equation}
    a[k]=\argmin_{a\in \Theta}\mathcal{C}_{H}(a,k).
\end{equation}
By learning the traffic tendency $a[k]\in \mathcal{A}$, the traffic tendency matrix $\mathbf{A}_a:\mathcal{A}\rightarrow \mathbb{R}^{\left(N-N_{out}\right)\times \left(N-N_{out}\right)}$ can be consistently updated and incorporated into modeling and control of traffic congestion.
\subsection{Traffic Feasibility Conditions}
Linear temporal logic (LTL) is used to specify the feasibility conditions of the conservation-based traffic coordination dynamics given in \eqref{MAIN} [\citep{wongpiromsarn2009receding}]. Every LTL formula consists of a set of atomic propositions, logical operators, and temporal operators. Logical operators include $\lnot$ (``negation''), $\vee$ (``disjunction''), $\wedge$ (``conjunction''), and $\Rightarrow$ (``implication''). Also,  $\square$ (``always''), $\bigcirc$ (``next''), $\lozenge$ (``eventually''), and $\mathcal{U}$ (``until'') are the temporal operators used in LTL formulas. 

Four traffic feasibility conditions are formally specified below to serve as formal constraints for optimal control definition as defined below. In particular, these feasibility conditions are used to determine admissible boundary inflow $\mathbf{u}[k]\in \mathbf{U}$. 

Traffic feasibility conditions are specified as follows:
\\
\textbf{Feasibility Condition 1:} Traffic density, defined as the number of cars at a road element, is a positive quantity everywhere in the NOIR. Also, it is assumed that every road element has  maximum capacity $\rho_{\mathrm{max}}$. Therefore, the number of cars cannot exceed $\rho_{\mathrm{max}}$ in every road element $i\in \mathcal{V}_I$. These two requirements can be formally specified as follows:
\begin{equation}
    \fbox{$\displaystyle\bigwedge_{i\in \mathcal{V}}\displaystyle\bigwedge_{h=0}^{N_\tau}\left(\left(\rho_i[k+h]\geq0\right)\wedge \left(\rho_i[k+h]\leq\rho_{\mathrm{max}}\right)\right)$}.\tag{$\Phi_1$}
\end{equation}
If feasibility condition $\Phi_1$ is satisfied at every road element, then, traffic over-saturation is avoided everywhere in the NOIR at discrete times $k$, $k+1$, $\cdots$, $k+N_\tau$.

\textbf{Feasibility Condition 2:}  Fraction $q_{j,i,a[k]}$ of the outflow $z_{i,a[k]}$ directed from $i\in \mathcal{V}$ toward $j \in \mathcal{O}_{i}$ must not exceed the available capacity of road element $i$ denoted by $C_j[k+h]=\rho_{\mathrm{max}}-\rho_j[k+h]$ at discrete times $k$, $k+1$, $\cdots$, $k+N_\tau$. This condition can be formally specified by
\begin{equation}
    \fbox{$\displaystyle\bigwedge_{i\in \mathcal{V}}\bigwedge_{j\in \mathcal{O}_{i}}\displaystyle\bigwedge_{h=0}^{N_\tau}\left(q_{_{j,i,a[k]}}z_{_{i,a[k]}}[k+h]\leq C_j[k+h]\right)$}.\tag{$\Phi_{2,a}$}
\end{equation}

\textbf{Feasibility Condition 3:} The inflow $y_{i,a[k]}$ must not exceed the available available capacity of road element $i\in \mathcal{V}_I$ denoted by $C_i[k+h]=\rho_{\mathrm{max}}-\rho_i[k+h]$ at discrete times $k$, $k+1$, $\cdots$, $k+N_\tau$. This requirement is formally specified by the following LTL formula:
\begin{equation}
    \fbox{$\displaystyle\bigwedge_{i\in \mathcal{V}}\displaystyle\bigwedge_{h=0}^{N_\tau}\left(y_{_{i,a[k]}}[k+h]\leq C_i[k+h]\right)$}\tag{$\Phi_{3,a}$}.
\end{equation}

\textbf{Feasibility Condition 4:} The boundary inflow needs to satisfy the following feasibility condition at discrete time $k$, $k+1$, $\cdots$, $k+N_\tau$:
\begin{equation}
    \fbox{$\displaystyle\bigwedge_{h=0}^{N_\tau}\left(\mathbf{u}[k+h]\in \mathbf{U}\right)$}.\tag{$\Phi_{4}$}
\end{equation}
While Feasibility Conditions $1$ through $4$ need to be satisfied at every discrete time $k$, the following ``optional'' condition is also implemented when the inflow demand is high:

\textbf{Optional Condition 5:} Boundary inflow needs to satisfy the following feasibility condition at discrete time $k$, $k+1$, $\cdots$, $k+N_\tau$:
\begin{equation}
\label{deltaineqequality}
    \fbox{$\displaystyle\bigwedge_{h=0}^{N_\tau}\sum_{i\in \mathcal{V}_{in}}\left({u}_i[k+h]=u_0\right)$}.\tag{$\Phi_{5}$}
\end{equation}
Boundary condition \eqref{deltaineqequality} constrains the number of vehicles entering
the NOIR to be exactly $u_0$ at any time $k$.{\color{black} Note that $u_0$ is an upper bound on vehicles entering the NOIR.  However, in the simulation results presented, traffic demand is significant such that the NOIR is maximally utilized by as many vehicles as possible.}
\\

 Given feasibility conditions $1$ through $4$, set 
 \begin{equation}
 \mathcal{P}=\{\left(\Phi_1,\Phi_{2,a},\Phi_{3,a},\Phi_4,\Phi_5\right)\big|a\in \mathcal{A}\}
 \end{equation}
 defines all possible atomic propositions. 
 \begin{figure}[ht]
\centering
\includegraphics[width=3.3  in]{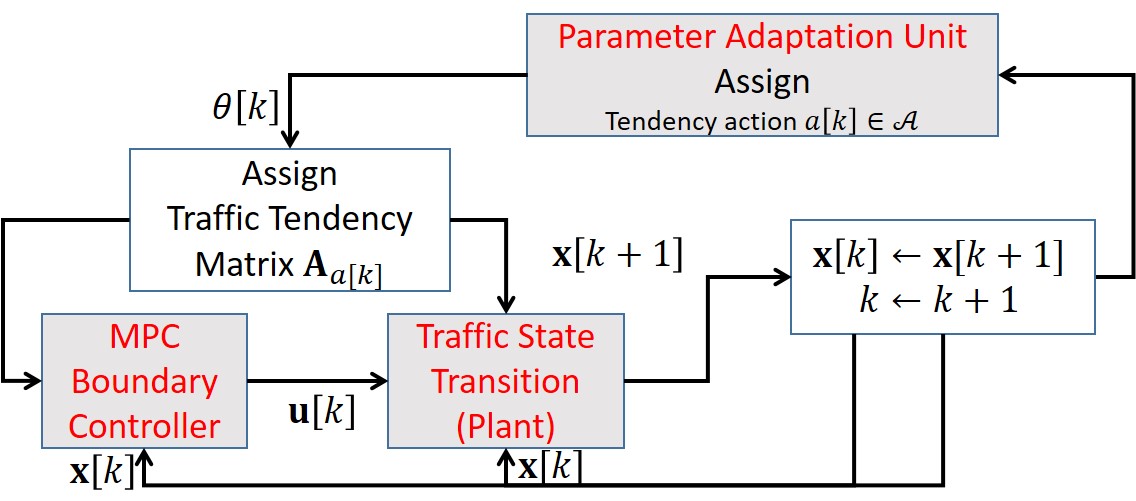}
\caption{Adaptive MPC traffic boundary controller.}
\label{MPOHILL}
\end{figure}
\section{Traffic Coordination Control}
\label{Traffic Coordination Control}
The boundary inflow $\mathbf{u}$ can be controlled by ramp meters situated at the inlet boundary nodes. 
Ramp meters apply an adaptive-MPC control design to determine the optimal boundary inflow $\mathbf{u}[k]$ so that the traffic congestion can be effectively and resiliently managed. The proposed adaptive-MPC design consists of three main components: (i) Plant (dynamics) (ii) Parameter Adaptation Unit (PAU), (iii) MPC Boundary Controller (See Fig. \ref{MPOHILL}). The traffic coordination dynamics obtained in \eqref{MAIN} is used to model the plant dynamics over the time interval $[t_k,t_{k+1}]$.  Given traffic tendency action $a[k]\in \mathcal{A}$, a PAU consistently learns and updates matrix $\mathbf{A}_{a[k]}$. Matrix $\mathbf{A}_{a[k]}$ is used by the MPC Boundary Controller to assign the input vector $\mathbf{u}[k]\in \mathbb{R}^{N_{in}}$ based on the updated plant dynamics.

The optimal boundary inflow $\mathbf{u}[k]=\mathbf{u}^*$ is determined by minimizing the $N_\tau$-step expected cost 
\begin{equation}
\label{RawCost}
\begin{split}
     \mathrm{C}=&\sum_{\tau=1}^{N_\tau}\left(\mathbf{x}^T[k+\tau]\mathbf{x}^T[k+\tau]+\beta\mathbf{u}^T[k+\tau]\mathbf{u}[k+\tau]\right),\\
\end{split}
\end{equation}
where $\mathrm{C}=\mathrm{C}\left(\mathbf{x}[k],\mathbf{u}[k+1],\cdots,\mathbf{u}[k+N_\tau]\right)$ scaling parameter $\beta>0$ is constant and all traffic feasibility conditions must be satisfied. The optimal control $\mathbf{u}[k]=\mathbf{u}^*$ is assigned as the solution of a  constrained quadratic programming problem that can be formally specified as follows:
\begin{equation}
\begin{split}
    \left(\mathbf{u}[k],\mathbf{u}[k+1],\cdots,\mathbf{u}[k+N_\tau]\right)=&\argmin\limits_{\left(\mathbf{u}[k+1],\cdots,\mathbf{u}[k+1]\right)\in \mathbf{U}^{N_\tau}} \mathrm{C}
    \\
    \mathrm{M}\models& \left(\Phi_1\wedge \Phi_4\wedge \Phi_5\right)\\
    \mathrm{M}\models& \left(\Phi_{2,a[k]}\wedge \Phi_{3,a[k]}\right)\\
\end{split}
.
\end{equation}
Note that the traffic tendency action $a[k]\in \mathcal{A}$ is learned based on empirical traffic information. Hence, action $a[k]\in \mathcal{A}$ is known  which in turn implies that cost function $\mathrm{C}$ can only be defined based on $\mathbf{x}[k],\mathbf{u}[k+1],\cdots,\mathbf{u}[k+N_\tau]$ at every discrete time $k$. Therefore, $\mathbf{u}^*=\mathbf{u}[k]$ can be assigned as the solution of the constrained quadratic programming problem at every discrete time $k$.
\begin{figure}[htb]
\centering
\includegraphics[width=3.3  in]{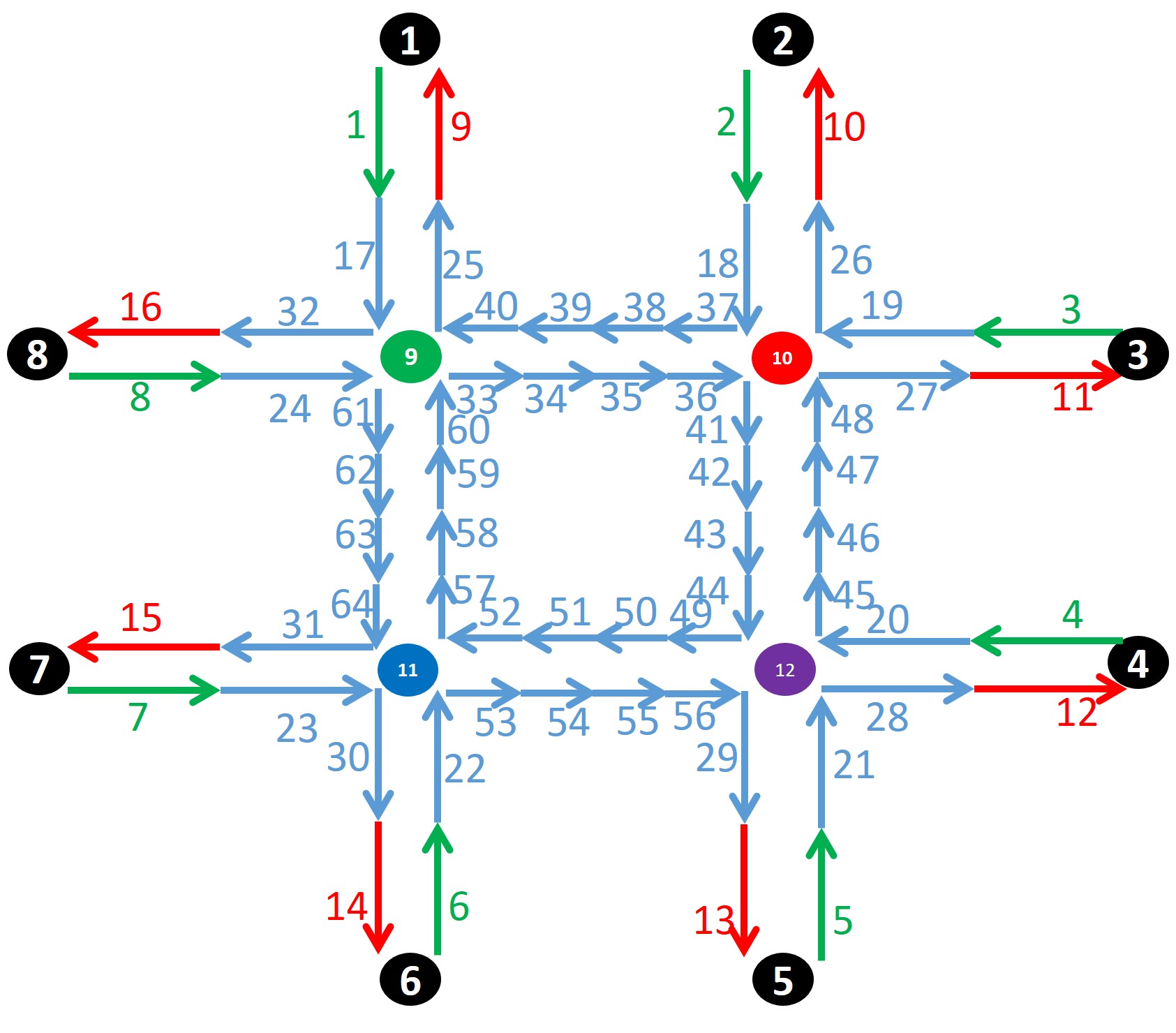}
\caption{Example NOIR with $24$ unidirectional roads. Every inlet or outlet road is filled by two serially-connected road elements while every interior road is filled by four road elements. }
\label{prelimexample}
\end{figure}

\begin{figure*}[htb]
\centering
\includegraphics[width=5.5  in]{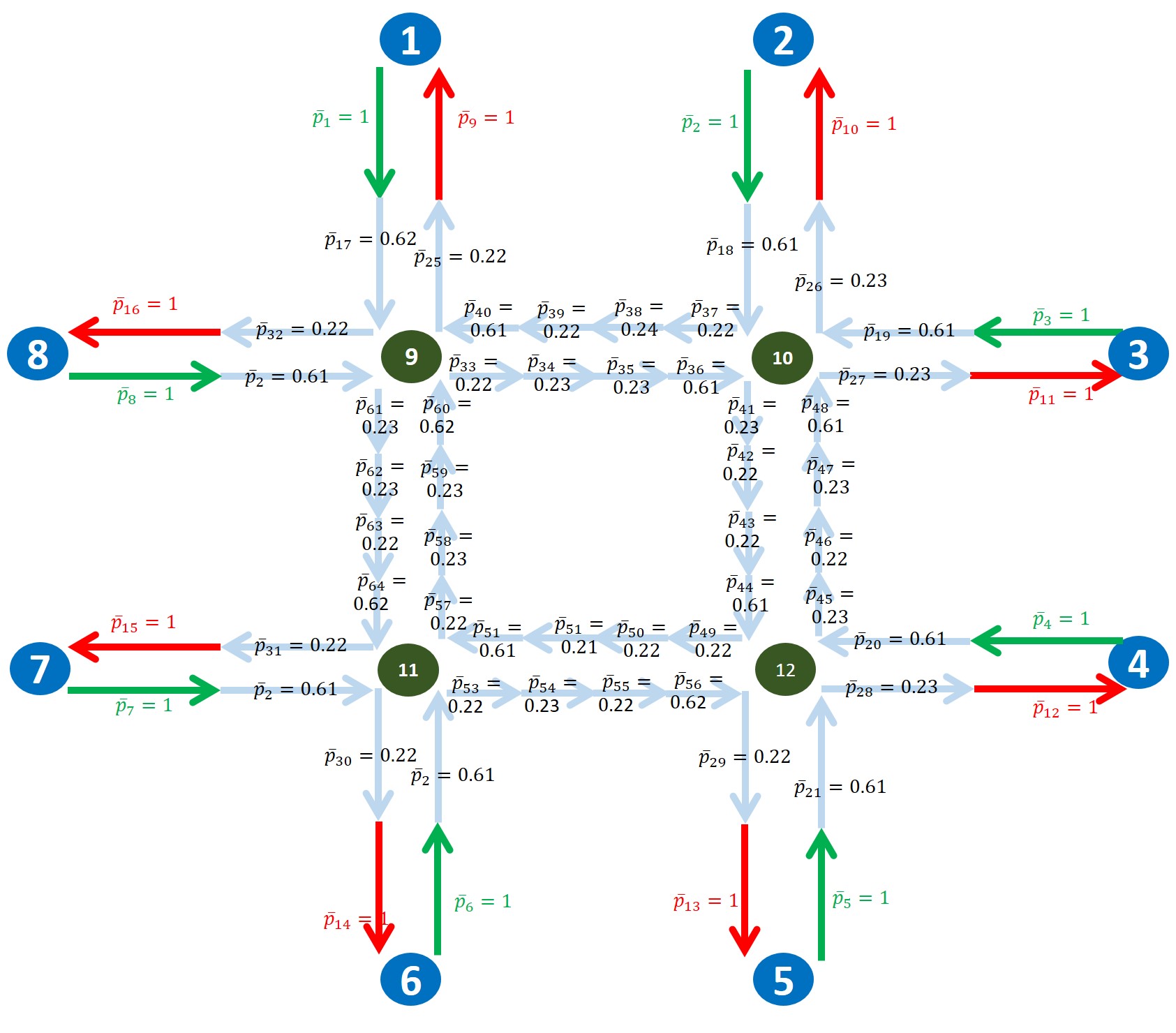}
\caption{Average tendency probability $\bar{p}_i$ at every node $i\in \mathcal{V}$. }
\label{mainsim2}
\end{figure*}


\begin{figure}[htb]
\centering
\includegraphics[width=3.3  in]{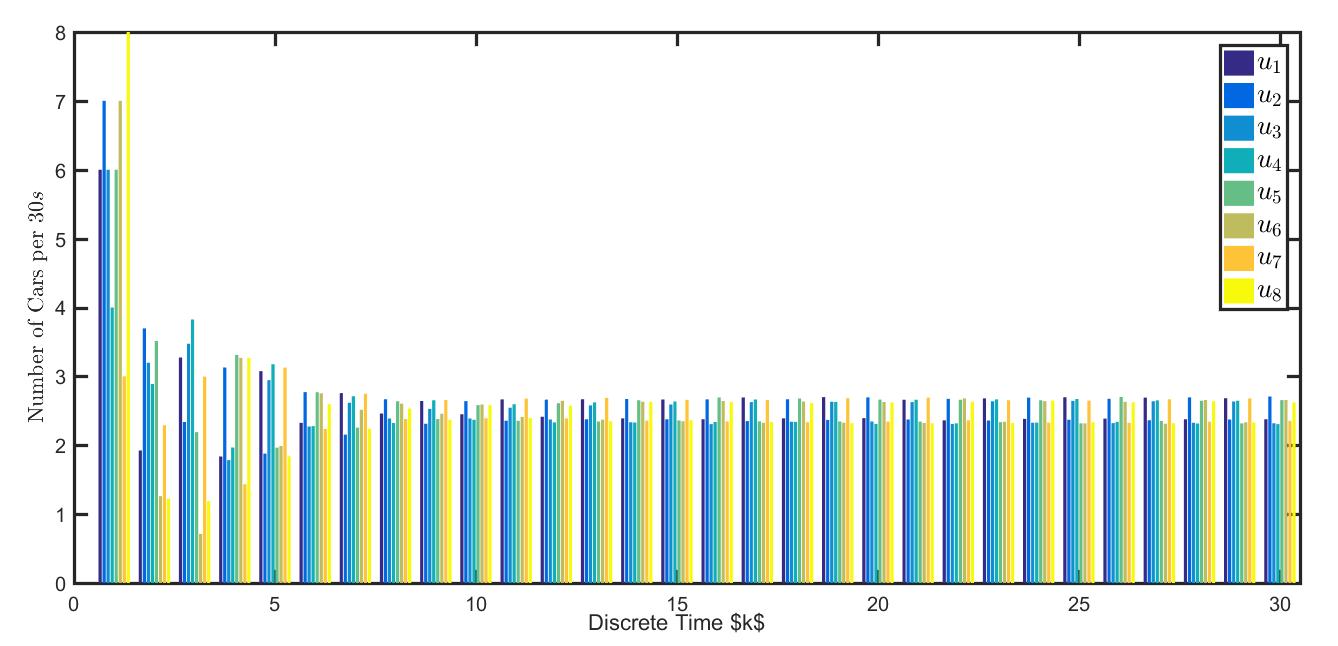}
\caption{Boundary control input $u_1[k]$ through $u_8[k]$ for $k=1,\cdots,30$}
\label{Input}
\end{figure}

\begin{figure}[htb]
\centering
\includegraphics[width=3.3  in]{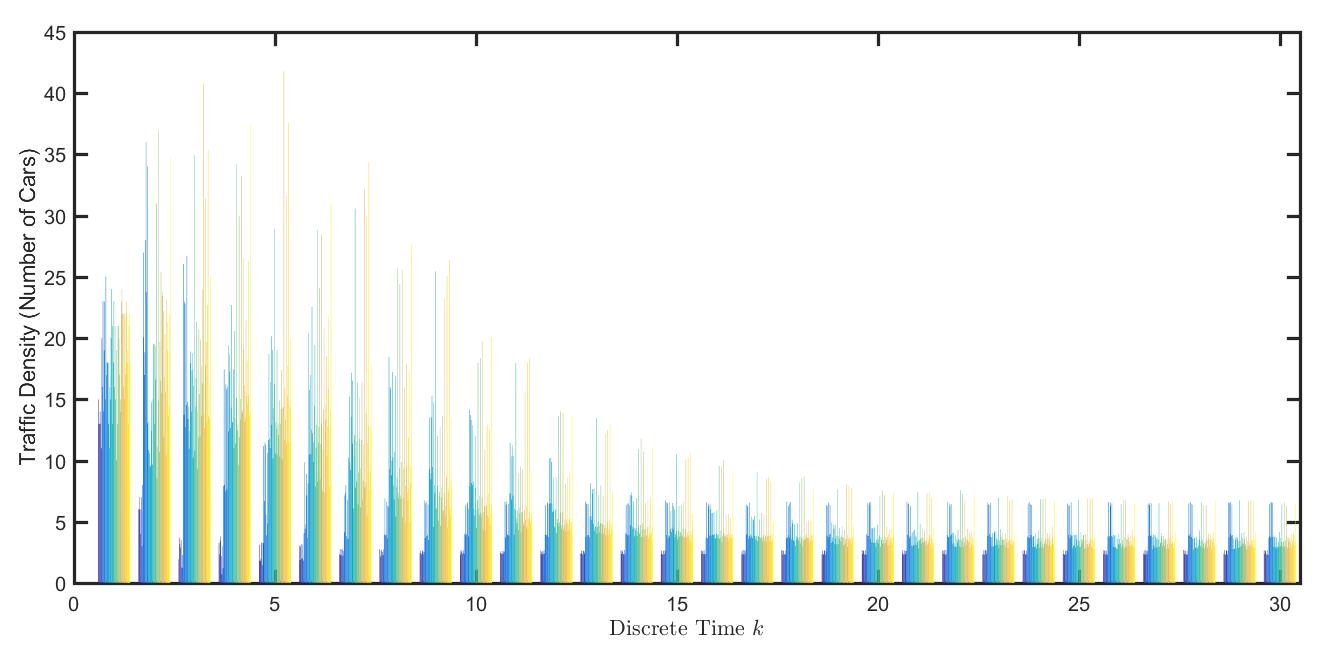}
\caption{Traffic density at interior road elements .}
\label{MPOHIL}
\end{figure}

\section{Simulation Results}\label{Simulation Results}
The NOIR consists of $24$ unidirectional roads where every interior road is filled out by four serially-connected road elements. Furthermore, every inlet or outlet road is filled out with two road elements. Communication between road elements is defined by graph $\mathcal{G}(\mathcal{V},\mathcal{E})$ where $\mathcal{V}=\{1,\cdots,64\}=\mathcal{V}_{in}\bigcup\mathcal{V}_{out}\bigcup\mathcal{V}_I$,  $\mathcal{V}_{in}=\{1,\cdots,8\}$,  $\mathcal{V}_{out}=\{9,\cdots,16\}$, and $\mathcal{V}_I=\{17,\cdots,64\}$ (See Fig. \ref{prelimexample}). We assume that the human intent is not changed, therefore, $\bar{p}_{i,a}=\bar{p}_i$ at every road element $i\in \mathcal{V}$ (See Fig. \ref{mainsim2}). We further assume that $u_0=20$, thus the number of vehicles entering the NOIR are restricted to be $20$ at any time $k$. Also, $\rho_{\mathrm{max}}=45$ is selected for the simulation.
Boundary control velocity inputs $u_1$ through $u_8$ are plotted versus discrete time $k$ for $k=1,\cdots,30$ in Fig. \ref{Input}. Fig. \ref{MPOHIL}  plots the  traffic density in all road elements versus discrete time $k$. It is seen that the number of vehicles is reduced as time goes ahead and the traffic congestion is successfully controlled. {\color{black}Figs. \ref{Input} and \ref{MPOHIL} imply that traffic density reaches the steady state values after about 20 time steps ($20\times 30=600s$)  in simulation while traffic consistently enters the NOIR through the inlet boundary nodes.}

\section{Conclusion}\label{Conclusion}
This paper offers a new data-driven  physics-inspired approach to effectively model and control traffic congestion. While existing physics-based approaches have commonly assigned traffic state (density and flow) by using the Fundamental Diagram. This paper suggests specifying the flow-density relation using empirical data. Therefore, the proposed approach offers several benefits: (i)
Traffic data is consistently incorporated, (ii) A high-fidelity
traffic coordination model is developed, (iii) Microscopic properties of a traffic system are Incorporated into planning, and (iv) Resilience of traffic congestion control is improved. Furthermore, feasibility conditions for traffic coordination in a large-scale urban network are formally specified using liner temporal logic.

\bibliography{ifacconf}             
                                                   







\end{document}